# Polarization-Tunable Colorimetric Metasurfaces for All-Optical and Non-Destructive Structural Characterization of Polymeric Nanofibers

Paula Kirya[1], Justin D. Hochberg[2], Han Sol Kim[2], Zaid Haddadin[3], Samantha Bordy[1], Jiuk Byun[4], Jonathan K. Pokorski[2,4], Lisa V. Poulikakos[1,4]

[1]Department of Mechanical and Aerospace Engineering, UC San Diego, La Jolla, CA, USA

[2] Aiiso Yufeng Li Family Department of Chemical and Nano Engineering, UC San Diego, La Jolla, CA, USA

[3]Department of Electrical and Computer Engineering, UC San Diego, La Jolla, CA, USA

[4]Program in Materials Science and Engineering, UC San Diego, La Jolla, CA, USA

**ABSTRACT**

Metasurfaces have pioneered significant improvements in sensing technology by tailoring strong optical responses to weak signals. When designed with anisotropic subwavelength geometries, metasurfaces can tune responses to varying polarization states of light. Leveraging this to quantify structural alignment in fibrous materials unveils an alternative to destructive characterization methods. This work introduces metasurface-enhanced polarized light microscopy (Meta-PoL), which employs polarization-tunable, guided-mode-resonant colorimetric metasurfaces to characterize molecular and bulk alignment of poly(ε-caprolactone) (PCL) nanofibers in a far-field configuration. PCL nanofibers drawn at 0%, 400%, and 900% ratios were interfaced with the studied metasurfaces. Metasurface resonances coinciding with the intrinsic drawn nanofiber resonances – confirmed by Stokes Polarimetry – produced the strongest colorimetric enhancement, resultant from alignment-specific nanofiber reflectivity. The enhancement degree corresponded with molecular and bulk alignments for each draw ratio, as measured through differential scanning calorimetry and scanning electron microscopy, respectively. Thus, Meta-PoL presents an all-optical, non-destructive, and quantitative measurement of nanofiber alignment.

# Introduction

Metasurfaces are optically thin surfaces composed of periodic subwavelength structures which manipulate light to engineer and enhance light-matter interactions, thus allowing them to introduce impactful improvements in sensing technology.[1,2] Tailoring the metasurface refractive index, geometry and periodicity gives rise to specific transmissive or reflective responses due to the distinct phase, amplitude, and/or polarization state from an incident optical signal.[2,3] When desirable, metasurfaces can be designed with anisotropic elemental geometry to generate polarization-sensitive responses.[3–7] This property has enabled metasurfaces to probe the structural properties of fibrous materials by selective enhancement of their interaction with polarized light. Recent studies have proposed leveraging metasurfaces and photonic surfaces to evaluate the organization of naturally occurring fibers, e.g. collagen, found in biological tissue microstructures of optically thin histological sections.[8–10] However, the characterization of natural or synthetic optically thick fibrous materials can also be achieved by probing the inherent optical effects of their anisotropic geometry at micro- and nanoscale dimensions.[9,11] This allows the evaluation of non-transparent samples, expanding how fibrous material characterization can be achieved with nanophotonics.

Nanofibers, polymeric fibers with dimensions in the nanoscale regime, have prominent applications in science, engineering, and medicine.[2] Polyesters are a class of polymers with robust mechanical properties, making such nanofibers optimal for biomedical and industrial applications. Poly(ε-caprolactone) (PCL) is a polyester commonly used in biomedical applications due to its slow degradation rate and biocompatibility, allowing its utilization for drug delivery and medical implants.[13,14] The development and production of these materials requires intensive characterization techniques to determine thermal, mechanical, and structural properties. Common methodologies include differential scanning calorimetry (DSC), tensile testing, and electron microscopy. While highly informational, these techniques are destructive in nature. Employing metasurfaces for all-optical structural characterization of synthetic fibers provides an advantageous, non-destructive alternative.

Here, we introduce metasurface-enhanced polarized light microscopy (Meta-PoL), wherein polarization-tunable, guided-mode-resonant colorimetric metasurfaces act as an all-optical, non-destructive, and contact-free platform to characterize the structural properties of PCL nanofibers. The PCL nanofibers were prepared by melt co-extrusion followed by uniaxial drawing, i.e. stretched past their original length, to induce controlled alterations in fiber alignment and morphology.[15] The studied metasurface gratings exhibit distinct color responses upon interaction with varying polarization states of light. PCL nanofibers of varying draw ratio (DR) – ratio of drawn fiber length to starting fiber length – were interfaced with the metasurfaces and characterized in a simple optical microscope to evaluate the distinguishability of their molecular and bulk alignment.[15]

Using the 1976 CIELAB color distance $\Delta E_{ab}$, we first identify the metasurface geometry which exhibits the strongest differential color response with and without the presence of the PCL nanofibers.[16] We then interface the best-performing metasurface with PCL nanofibers of varying draw ratios and quantify a clear enhancement in reflection intensity within the resonant wavelengths of the metasurface, corresponding to the degree of nanofiber alignment. This effect is attributed to the varying degrees of elliptical polarization of the light upon interaction with PCL nanofibers with draw ratios that are more favorable for fiber alignment, as we demonstrate with Stokes Polarimetry. The results presented here introduce and experimentally demonstrate polarization-tunable metasurfaces as a quantitative imaging platform to non-destructively characterize the structural properties of polymeric nanofibers by means of the metasurface colorimetric response, paving the way toward versatile societal and industrial characterization platforms.

## Results and Discussion

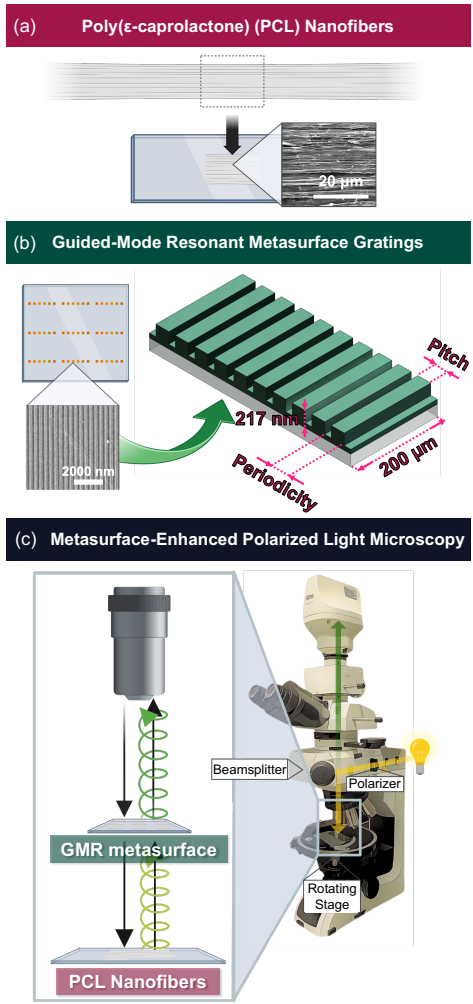

*Figure 1. Schematic of metasurface-enhanced polarized light microscopy (Meta-PoL) platform. (a) The investigated PCL nanofibers were melt-extruded and were drawn to 0% (DR1), 400% (DR5), or 900% (DR 10) uniaxial strain. Sections of these fibers were excised and taped onto a glass slide for imaging. (b) The silicon nitride GMR metasurfaces with grating pitches 195, 185, 175, 160, 150, and 140 nm were fabricated onto a silicon dioxide wafer in a 3x3 array. (c) The PCL nanofibers were placed underneath the GMR metasurfaces, separated by a 2.52±0.16 mm thick spacer (spacer thickness averaged across 5 points of measurement). The samples were imaged under episcopic illumination with the Nikon ECLIPSE LV100ND polarized light microscope. Illumination from a halogen lamp was sent through a linear polarizer and passed through a beamsplitter to illuminate the GMR metasurface + PCL nanofiber sample. Inset: Linearly polarized light obtains a degree of ellipticity when the light is reflected off the PCL nanofibers, which is further altered upon transmission back through the metasurface. The resulting reflected color of the GMR metasurface is interpreted with CIE Colorimetry. Graphic of Guided-Mode Resonant Gratings created in SolidWorks. Created in BioRender. Poulikakos, L. (2025) https://BioRender.com/w05j600.*

Metasurface-enhanced polarized light microscopy (Meta-PoL), our experimental characterization technique, is shown schematically in Figure 1. PCL nanofibers (Figure 1a) are imaged with guided-mode resonant (GMR) metasurfaces (Figure 1b) with linearly polarized illumination in a polarized light microscope. The metasurfaces are separated from PCL nanofibers with a spacer for contact-free imaging (see Supporting Information, Figure S2, S7). Figure 1c illustrates our metasurface-enabled characterization methodology. Based on the physics of the GMR metasurface, a portion of the incident light is reflected, yielding a distinct, high-quality-factor structural color response. The incident light transmitted through the GMR metasurface interacts directly with the nanofibers. Depending on their optical anisotropy, the nanofibers will reflect light with varying degrees of ellipticity which again transmits back through the GMR metasurface. This back-reflected light alters the final color and intensity response produced by the metasurface. The metasurface optical axis is oriented parallel to the incident linearly polarized light, while the optical axis of the nanofiber sample is oriented 45° relative to the metasurface optical axis to maximize reflected light ellipticity. The high sensitivity of the metasurface optical response to varying polarization states of light thus enables a quantitative classification of nanofiber structural properties.

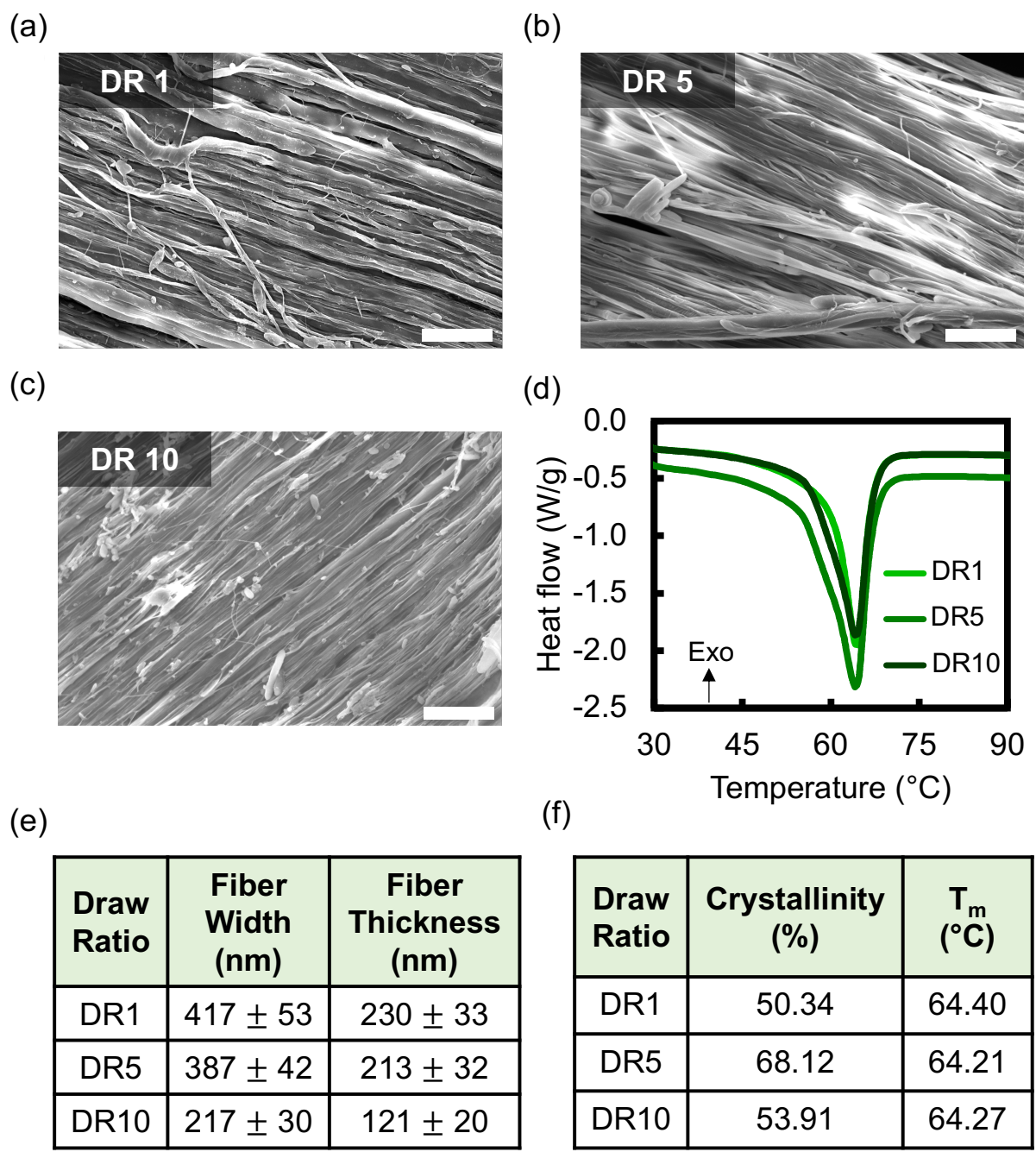

*Figure 2.* Characterization of PCL nanofibers of varying draw ratio. SEM image of (a) DR1, (b) DR5, and (c) DR10. Scale bars for a-c are 10 μm. (d) First heating run DSC thermograms of DR1, DR5, and DR10. (e) Mean and standard deviation for the thickness and width of DR1, DR5, and

*DR10 analyzed from SEM images measured with ImageJ. (f) Crystallinity and melting temperature of DR1, DR5 and DR10 PCL nanofibers calculated from the first DSC heating run.*

The GMR metasurfaces studied in this work are shown schematically in Figure 1b, where a 217 nm-thick layer of $Si_3N_4$ was deposited on a $SiO_2$ substrate. Nanoscale gratings with varying grating pitches [140 nm, 150 nm, 160 nm, 175 nm, 185 nm, 195 nm] and etch depths [57.68±0.21 nm, 56.11±3.01 nm, 47.32±0.90 nm, 41.15±0.3141 nm, 40.08±0.68 nm, 42.75±0.97 nm] were studied (see Supporting Information, Figures S10-S11). An example GMR metasurface with a 210 nm grating pitch is shown in a scanning electron micrograph (SEM) in Figure 1b.

PCL nanofibers are prepared as has previously been demonstrated (see Supporting Information, S5 and Figure S12).[17–19] Following fiber isolation, the nanofibers are then drawn to various draw ratios (see Supporting Information, S6 and Figure S13). PCL nanofiber length is controllably increased with applied strain, where a strain of 0%, 400% and 900% are denoted as Draw Ratio 1 (DR1), Draw Ratio 5 (DR5), and Draw Ratio 10 (DR10). PCL nanofibers at these draw ratios were studied in this work. Note that the nanofiber samples are optically thick, i.e. most of the incident light will be reflected off the sample (see Supporting Information, Figure S4). SEM images of each PCL nanofiber draw ratio are shown in Figure 2a-c, while differential scanning calorimetry (DSC) results and additional characterization for each sample is shown in Figure 2d-f. Fiber width and thickness were determined by analyzing SEM images with ImageJ software (see Supporting Information, S8). DR1, DR5 and DR10 showed 417, 387 and 217 nm widths, respectively, and 230, 213 and 121 nm thicknesses, respectively. Uniaxial drawing of PCL nanofibers resulted in a significant decrease of both fiber width and thickness, yielding ~400% thinner fibers upon the application of 900% strain.

Percent crystallinity and melting point of the PCL nanofibers was obtained from the first heating endotherm of DSC analysis (see Supporting Information, S7). DR1, DR5 and DR10 exhibited similar melting points ($T_m$) at ~64 °C. 139 J/g was used as the standard heat of fusion of PCL for calculating percent crystallinity.[20] DR5 (68.12%) showed higher crystallinity compared to DR1 (50.34%), explainable by uniaxial drawing improving molecular-level orientation of PCL chains within the nanofibers.[21] Above DR7, nanofiber structural integrity reduced. Thus, the DR10 sample displays lower crystallinity (53.91%) when compared with DR5 nanofibers (68.12%) due to this physical breakage, and the subsequent loss of enthalpic change otherwise available to the system.

Further analysis with OrientationJ (an ImageJ/FIJI plug-in) measured decreased bulk fiber alignment with increasing draw ratio (see Supporting Information, S8). This was determined from the mean coherency (i.e. degree of fiber alignment) from 10 SEM images per draw ratio. Coherency decreased from DR1 (0.72±0.05, CV = 7.45%) and DR5 (0.71±0.08, CV = 10.99%) to DR10 (0.55±0.17, CV = 30.68%). Thus, these evaluations confirm that increasing draw ratio diminishes fiber dimensions and bulk fiber alignment.

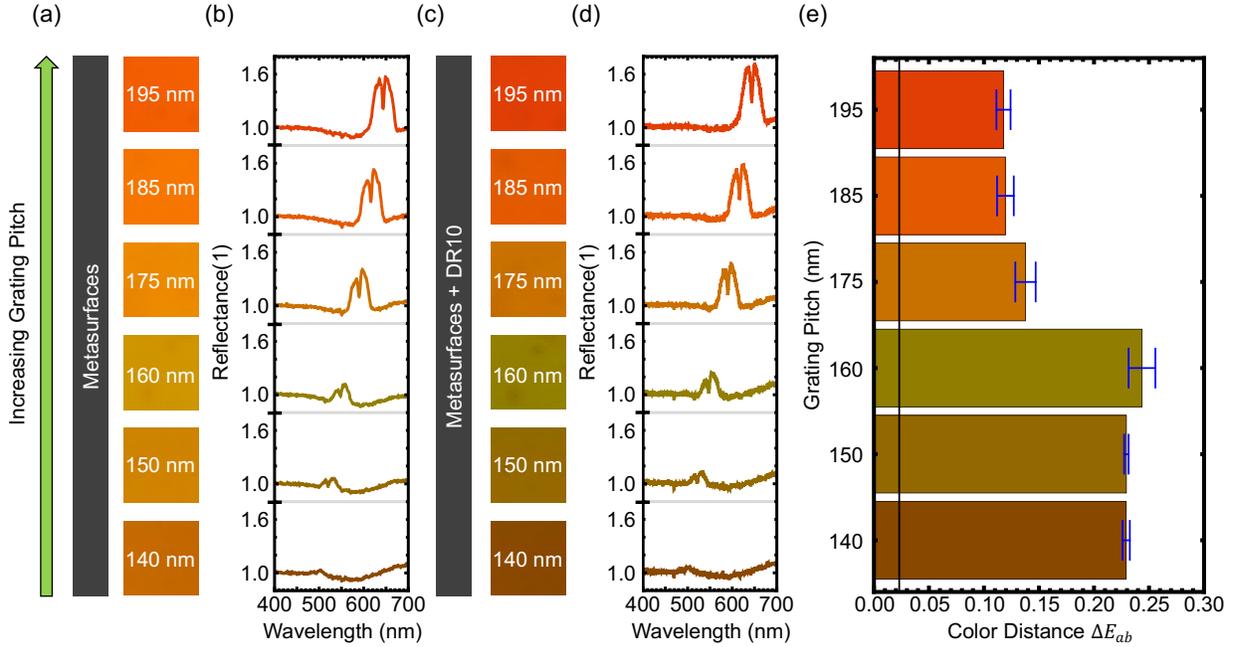

*Figure 3. Optical micrographs (a,c) and reflectance spectra (b,d) of GMR metasurfaces oriented at 0° along the horizontal without (parts a,b) and with (parts c,d) the presence of DR10 PCL nanofibers oriented at 45° with respect to the horizontal. Measurements are recorded under incident vertically polarized light. (e) Histogram of the average color distance $\Delta E_{ab}$ comparing the 1976 CIELAB color values of each metasurface with and without the presence of DR10 PCL nanofibers. Histogram colors represent the RGB color of the metasurface measured in the presence of the DR10 PCL nanofibers. Error bars correspond to the standard deviation of $\Delta E_{ab}$ for the three sets of images. The just noticeable difference (JND) which can be detected by the human eye (JND ≈ 0.023) is denoted by the vertical black line.*

Figure 3a,c shows the optical microscope images of our GMR metasurfaces under vertically polarized incident light, which were acquired at 5x magnification when no PCL nanofibers are present (Figure 3a) and when interfaced with DR10 PCL nanofibers (Figure 3c). The reflectance spectra of these metasurfaces without (Figure 3b) and with (Figure 3d) the PCL nanofibers were also acquired (see Supporting Information, S2). In principle, increasing periodicity – equivalent to twice the grating pitch for the studied metasurfaces – redshifts the resonant wavelength.[5,6,22,23] . Additionally, pronounced sidewall angles were observed at finer grating pitches (see Supporting Information, S4), decreasing electric field enhancement and consequentially the resonance reflection amplitude.[23,24] The narrow-bandwidth resonances in the metasurface reflectance is characteristic of guided-mode resonances, enabling high-sensitivity colorimetric analysis that would be challenging with alternative broad-bandwidth colorimetric metasurfaces fabricated e.g. from plasmonic metals[25–29] or high-refractive-index dielectrics which exhibit significant losses at visible frequencies.[30–32]

The observed colorimetric response with the PCL nanofibers was quantified using the 1976 CIELAB color space. In this color space, for two colors with chromaticity coordinates $(L^*_1, a^*_1, b^*_1)$ and $(L^*_2, a^*_2, b^*_2)$, we can define the color distance $\Delta E_{ab} = \sqrt{(L^*_2 - L^*_1)^2 + (a^*_2 - a^*_1)^2 + (b^*_2 - b^*_1)^2}$ which quantifies the separation between two colors and has been applied as a rough guide to their discriminability by the human eye.[16,33,34] $\Delta E_{ab}$ ranges from [0,1], where the just noticeable difference (JND) detectable by the human eye has been determined as $JND \approx 0.023$.[34] Figure 3c shows the color distance $\Delta E_{ab}$ for each metasurface grating pitch studied, with and without DR10 PCL nanofibers. The metasurfaces with a 160 nm grating pitch presented the largest color distance ($\Delta E_{ab} = 0.2436 \pm 0.01217$). While the strongest color distance ($\Delta E_{ab}$) was observed when interfaced with DR10, this metasurface presented the largest color distance across all draw ratios (see Supporting Information, Figure S3, Table S1). This best-performing metasurface will now be further investigated to comparatively assess nanofibers of varying draw ratios.

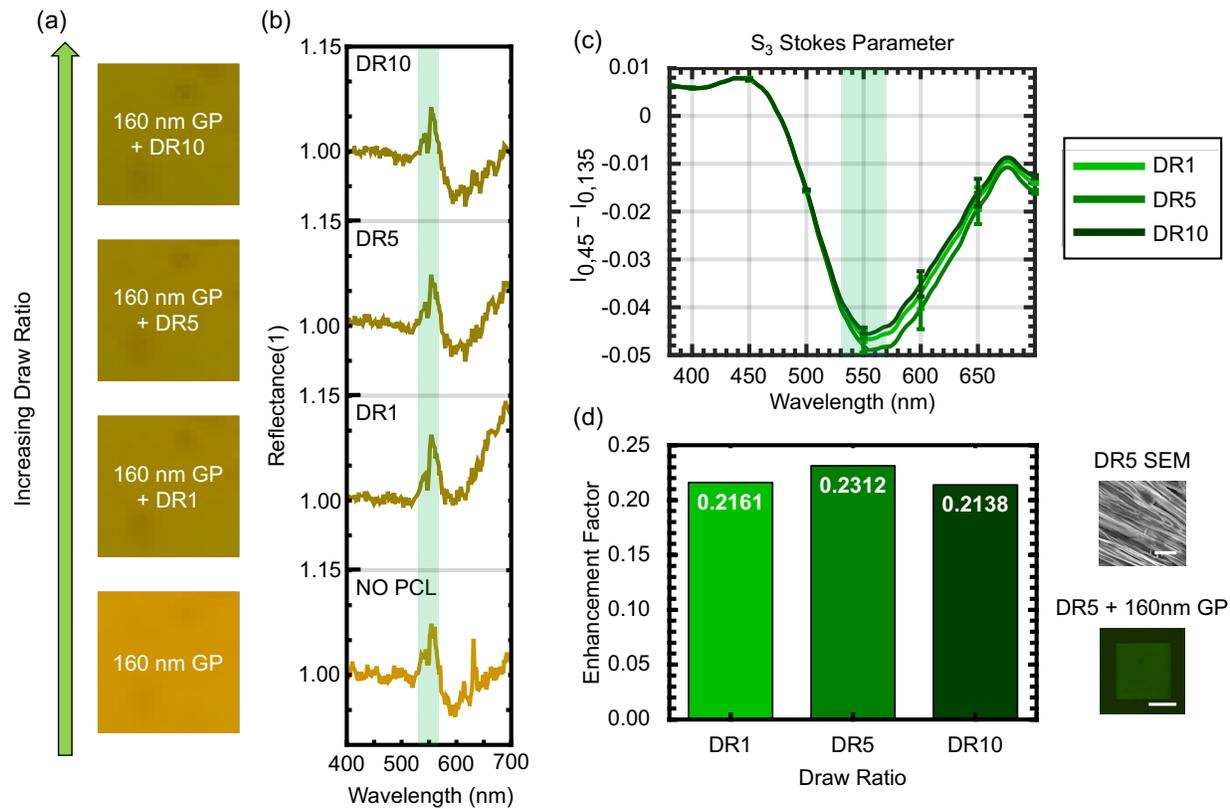

*Figure 4.* Optical micrographs (part a) and reflectance spectra (part b) of 160 nm grating pitch GMR metasurfaces oriented parallel to the incident linearly polarized light, without and with the presence of DR 1, 5, and 10 PCL nanofibers, which were oriented 45° relative to the metasurface optical axis. Reflectance spectra were taken over four distinct regions of the drawn PCL nanofibers interfaced with the 160 nm grating pitch GMR metasurface. These spectra were averaged and smoothed using a Gaussian-weighted moving average filter with a 10-element window in MATLAB. Metasurface spectra of the 160 nm grating pitch metasurface without the PCL nanofibers was taken with the metasurface over a glass slide separated by a 2.52±0.16 mm thick spacer to represent the imaging process. All spectra were normalized by the spectrum of the silicon nitride substrate placed over a glass slide with the spacer. (c) Reflectance spectra corresponding to the $S_3$ Stokes parameter for DR1, DR5, and DR10 PCL nanofibers. Acquired $S_3$
13

*spectra for 4 different regions of DR1, DR5, and DR10 were averaged and smoothed with a Gaussian-weighted moving average filter with a 250-datapoint window in MATLAB. All spectra were normalized by the spectrum of a ThorLabs Protected Silver Mirror (PF10"03-P01 ⌀1") at the same incident polarization and intensity. (d) Histogram of the metasurface enhancement factor $\left(\frac{I_{PCL}-I_{META}}{I_{META}}\right)$ of the 160 nm grating pitch GMR metasurface interfaced with PCL nanofibers of DR1, DR5, and DR10, respectively, illuminated by 550±20 nm light via a notch filter. Right top: SEM of DR5 PCL nanofibers. Right bottom: Optical micrograph of metasurface interfaced with DR5 PCL nanofibers illuminated by the filtered light. Mean measured intensity calculated over the four quadrants of the drawn PCL nanofibers was determined with NIS-Elements Basic Research ROI statistics, in which a square 189.6x189.6 µm ROI was drawn over the metasurface area for intensity measurement. Substrate containing the metasurfaces was vertically separated from the PCL nanofibers by a 2.98 mm thick spacer. Scale bars for DR5 SEM and DR5 + 160 nm GP optical micrograph are 10 µm and 100 µm, respectively.*

While increasing the PCL nanofiber draw ratio alters the molecular and bulk nanofiber alignment, this can be challenging to verify with existing methodologies such as DSC or SEM.[35–37] Figure 4 demonstrates the ability of the studied metasurfaces to optically discern small changes in nanofiber alignment with varying draw ratio. We interfaced the 160 nm grating pitch metasurface, which showed the strongest response in Figure 3, with DR1, DR5 and DR10 PCL nanofibers, where example SEM images are shown in Figure 2.

Figure 4a,b shows optical micrographs and reflectance spectra, respectively, of the 160 nm GP metasurface on its own and interfaced with nanofibers of increasing draw ratio (from bottom to top). The narrow-bandwidth character of the GMR metasurface reflectance resonance enables the detection of even slight changes in the light interacting with the metasurface, as seen in the clear spectral differences shown in Figure 4b. When linearly polarized light reflects off an optically anisotropic medium such as the PCL nanofibers, the reflected light will obtain a degree of ellipticity. We hypothesize that increasing the PCL nanofiber draw ratio alters the degree of ellipticity of the reflected light.

This small variation in ellipticity is difficult to measure with conventional polarized light microscopy. However, the reflective response of our metasurface gratings exhibits high sensitivity to small changes in the polarization state of light, allowing nanofiber draw ratios to be optically distinguished. To further quantify the ellipticity of the light reflected from the nanofibers, we performed Stokes polarimetry measurements (see Supporting Information, S2). The Stokes parameters describe the polarization state of light. There are four recognized Stokes parameters ($S_0$-$S_3$), with $S_3$ describing the right-handedness ($S_3 > 0$), left-handedness ($S_3 < 0$), or absence of handedness ($S_3 = 0$) of the light. This is given by the equation $S_3 = 2E_{0x}E_{0y}\sin\boldsymbol{\delta}$, where $E_{0x}$ and $E_{0y}$ are the amplitude of the electric field in the x and y direction, respectively, and $\boldsymbol{\delta}$ is the relative phase difference between the electric field on the x-axis and the electric field on the y-axis.[38] Figure 4c shows the Stokes parameter ($S_3$) for DR1, DR5 and DR10 nanofibers upon linearly polarized light excitation. The significant ellipticity observed in the light reflected from the nanofibers demonstrates that the nanofibers exhibit optical anisotropy, enabling a conversion of linear to elliptically polarized light. Moreover, the resonant wavelengths of the 160 nm grating pitch metasurface fall within the wavelengths of highest ellipticity (mean lower and upper full-width half maximum bounds for $S_3$: $\bar{\lambda}_{lower} = 509.16 \pm 0.473$ nm, $\bar{\lambda}_{upper} = 635.52 \pm 1.522$ nm), thus explaining its strong reflective response (see Supporting Information, Figures S5, S6).

Increased metasurface reflectivity produced by the nanofibers was quantified with an enhancement factor. The 160 nm grating pitch metasurface interfaced with a drawn PCL nanofiber (DR1, DR5, or DR10) was imaged under the illumination of 550±20 nm light via a notch filter to isolate the resonant wavelengths of this metasurface and observe the resulting enhancement. Figure 4d displays a histogram of the enhancement factor for DR1, DR5, and DR10 PCL nanofibers. The metasurface enhancement factor was calculated with the equation $\frac{I_{PCL+META} - I_{META}}{I_{META}}$, where $I_{PCL+META}$ is the mean pixel intensity for images captured of the metasurface interfaced with the drawn nanofiber and $I_{META}$ is the intensity of the metasurface by itself. Here, we see the greatest enhancement with DR5 (0.2312), followed by DR1 (0.2161) then DR10 (0.2138).

Crystallinity increased significantly from DR1 (50.3%) to DR5 (68.12%) and dropped significantly with DR10 (53.9%) (Figure 2d,f). The drawing process aligns macromolecules within the polymer, increasing crystallinity, light scattering, and birefringence on the molecular scale.[39,40] Thus, we can expect stronger visible light scattering and elliptical polarization with more crystalline nanofibers. This may explain why DR5 demonstrates the strongest reflection intensity (0.1433±0.3166 × $10^{-2}$), ellipticity ($|I_{0,45} - I_{0,135}|_{max_{DR5}} = 0.0490$), and enhancement factor (0.2312).

Furthermore, nanofiber thinness promotes light transmission.[41] The thin DR10 nanofibers (121±20 nm thickness) transmit substantially more light than the DR1 (230±33 nm thickness) and DR5 (213±32 nm thickness) nanofibers. DR10 nanofibers are also highly disordered compared to DR1 and DR5, as demonstrated with the coherency (DR1 = 0.72±0.05, DR5 = 0.71±0.08, DR10 = 0.55±0.17) (see Supporting Information, S8, Table S2). Thus, high disorder in DR10 may drive multiple scattering events, producing a comparatively low reflection intensity (DR10 = $0.1412 \pm 0.3581 \times 10^{-2}$, DR1 = $0.1415 \pm 0.7933 \times 10^{-3}$), ellipticity ($|I_{0,45} - I_{0,135}|_{max_{DR10}} = 0.0456$, $|I_{0,45} - I_{0,135}|_{max_{DR1}} = 0.0468$) and enhancement factor (DR10 = 0.2138, DR1 = 0.2161).

Our results demonstrate that Meta-PoL can leverage GMR metasurfaces to provide an all-optical, non-destructive methodology to detect structural properties of optically thick PCL nanofiber samples with high sensitivity, surpassing that of established optical methods, by mapping the molecular and bulk structural properties of the sample onto a quantitative color response in a simple polarized light microscope.

## Conclusion and Outlook

In summary, we have introduced metasurface-enhanced polarized light microscopy (Meta-PoL) and demonstrated that polarization-tunable guided-mode-resonant metasurfaces are capable of quantifying alignment in optically thick, linearly birefringent polymeric nanofiber samples with reflected structural color. The bulk and molecular anisotropy within the studied nanofibers induces optical anisotropy specific to the degree of fiber alignment, thus offering a novel non-destructive approach to material characterization. Nanofiber crystallinity increased up to DR5, then lowered with DR10 due to structural damage. The physical breakdown of DR10 also introduced high disorder on the bulk scale compared to DR1 and DR5. Thus, there is a clear draw ratio threshold for optimal PCL nanofiber alignment.

Experimental Stokes Parameter measurements showed that nanofiber anisotropy alters the polarization state of incident polarized light to an elliptical polarization in the reflected optical signal, corresponding to PCL nanofiber alignment. The spectral regime of this ellipticity agreed with the guided-mode-resonance of the 160 nm grating pitch metasurface, which presented the greatest color distance when compared to other metasurface geometries. Hence, when interfaced with the drawn PCL nanofibers, ellipticity of the reflected light was correlated with the change in reflected intensity of the light emerging from the metasurface. This was quantified with the enhancement factor $\left(\frac{I_{PCL+META}-I_{META}}{I_{META}}\right)$, which was measured by recording the mean pixel intensity for images of the 160 nm grating pitch GMR metasurface interfaced with PCL nanofibers of DR1 (0.2161), DR5 (0.2312), and DR10 (0.2138), respectively, illuminated by 550±40 nm light. With this measurement we observed the highest intensity enhancement with DR5, which was highly aligned on both molecular and bulk scales, followed by DR1 which exhibited the highest alignment in the bulk scale, and then DR10. DR10 was significantly thinner and more disordered than DR1 and DR5, indicating that drawing to 900% strain reduces the structural integrity of the material.

Given the promising characterization capabilities of the GMR metasurfaces utilized in this study, metasurface sensitivity may be improved with fabrication methods more favorable for finer grating pitches. Furthermore, more discriminative optical responses can be achieved by optimizing metasurface geometry with inverse design methods.[42] The presented Meta-PoL methodology holds sufficient potential for analogous or more pronounced fiber alignment sensing with other synthetic or biological samples as this technique leverages structural anisotropy. Thus, Meta-PoL can extend to various fibrous materials used across multiple industries.

ASSOCIATED CONTENT

**Supporting Information**.

The Supporting Information is available free of charge. Detailed explanation of experimental methodology and additional data are provided (PDF).

AUTHOR INFORMATION

**Corresponding Author**

*L.V.P. University of California San Diego, La Jolla, CA, lpoulikakos@ucsd.edu

**Author Contributions**

Following CRediT definitions: conceptualization: P.K., L.V.P., J.D.H., J.K.P.; data curation: P.K. H.S.K., J.D.H., Z.H.; formal analysis: P.K., H.S.K., Z.H.; funding acquisition: L.V.P., J.K.P.; investigation: P.K., J.D.H., H.S.K., S.B., J.B.; methodology: P.K., J.D.H, H.S.K.; project administration: P.K., resources: L.V.P., J.K.P.; supervision: L.V.P., J.K.P.; validation: P.K., H.S.K.; visualization: P.K., H.S.K; writing – original draft: P.K.; writing – reviewing and editing: P.K., L.V.P., H.S.K., J.D.H., Z.H. All authors have given approval to the final version of the manuscript.

**Funding Sources**

L.V.P., Z.H., and P.K. gratefully acknowledge funding from the Beckman Young Investigator award by the Arnold and Mabel Beckman Foundation (Project Number: 30155266). P.K. acknowledges funding from the Optica Amplify Scholarship, the Optica Women's Scholarship, and the Selected Professions Fellowship from the American Association of University Women (AAUW) and the National Science Foundation Graduate Research Fellowship Program (Grant No.

DGE-2038238). P.K., J.D.H., S.B., J.B., J.KP., and L.V.P. acknowledge funding support from the UC San Diego Materials Research Science and Engineering Center (UCSD MRSEC), supported by the National Science Foundation (Grant DMR-2011924). H.S.K. acknowledges funding support from the US Department of Energy (# DE-EE0009296). This work was performed in part at the San Diego Nanotechnology Infrastructure (SDNI) of UCSD, a member of the National Nanotechnology Coordinated Infrastructure (NNCI), which is supported by the National Science Foundation (Grant ECCS-2025752).

**Notes**

The authors declare no competing financial interest.

**ACKNOWLEDGMENT**

The authors thank Applied Materials (Robert Visser, Nir Yahav, Zihao Yang, David Sell, Adi de la Zerda) for providing silicon nitride wafers for metasurface fabrication. P.K. thanks Shahrose Khan for assistance with color distance extraction code. This material is based upon work supported by the National Science Foundation Graduate Research Fellowship Program under Grant No. DGE-2038238. Any opinions, findings, and conclusions or recommendations expressed in this material are those of the author(s) and do not necessarily reflect the views of the National Science Foundation.

REFERENCES

(1) Tseng, M. L.; Jahani, Y.; Leitis, A.; Altug, H. Dielectric Metasurfaces Enabling Advanced Optical Biosensors. *ACS Photonics* **2021**, *8* (1), 47–60. https://doi.org/10.1021/acsphotonics.0c01030.
(2) Chen, H.-T.; Taylor, A. J.; Yu, N. A Review of Metasurfaces: Physics and Applications. *Rep. Prog. Phys.* **2016**, *79* (7), 076401. https://doi.org/10.1088/0034-4885/79/7/076401.

(3) Bukhari, S. S.; Vardaxoglou, J.; Whittow, W. A Metasurfaces Review: Definitions and Applications. *Appl. Sci.* **2019**, *9* (13), 2727.
(4) Chen, H.-T.; Taylor, A. J.; Yu, N. A Review of Metasurfaces: Physics and Applications. *Rep. Prog. Phys.* **2016**, *79* (7), 076401.
(5) Uddin, M. J.; Magnusson, R. Highly Efficient Color Filter Array Using Resonant $Si_3N_4$ Gratings. *Opt. Express* **2013**, *21* (10), 12495. https://doi.org/10.1364/OE.21.012495.
(6) Uddin, M. J.; Khaleque, T.; Magnusson, R. Guided-Mode Resonant Polarization-Controlled Tunable Color Filters. *Opt. Express* **2014**, *22* (10), 12307–12315. https://doi.org/10.1364/OE.22.012307.
(7) Haddadin, Z.; Khan, S.; Poulikakos, L. V. Cutting Corners to Suppress High-Order Modes in Mie Resonator Arrays. *ACS Photonics* **2024**, *11* (1), 187–195. https://doi.org/10.1021/acsphotonics.3c01270.
(8) Haddadin, Z.; Pike, T.; Moses, J. J.; Poulikakos, L. V. Colorimetric Metasurfaces Shed Light on Fibrous Biological Tissue. *J. Mater. Chem. C* **2021**, *9* (35), 11619–11639.
(9) Poulikakos, L. V.; Lawrence, M.; Barton, D. R.; Jeffrey, S. S.; Dionne, J. A. Guided-Mode-Resonant Dielectric Metasurfaces for Colorimetric Imaging of Material Anisotropy in Fibrous Biological Tissue. *ACS Photonics* **2020**, *7* (11), 3216–3227. https://doi.org/10.1021/acsphotonics.0c01303.
(10) Kirya, P.; Mestre-Farrera, A.; Yang, J.; Poulikakos, L. V. Leveraging Optical Anisotropy of the Morpho Butterfly Wing for Quantitative, Stain-Free, and Contact-Free Assessment of Biological Tissue Microstructures. *Adv. Mater.* **2025**, *37* (12), 2407728. https://doi.org/10.1002/adma.202407728.
(11) Tuchin, V. V.; Wang, L. V.; Zimnyakov, D. A. *Optical Polarization in Biomedical Applications*; Biological and medical physics, biomedical engineering; Springer: Berlin ; New York, 2006.
(12) Lou, L.; Osemwegie, O.; Ramkumar, S. S. Functional Nanofibers and Their Applications. *Ind. Eng. Chem. Res.* **2020**, *59* (13), 5439–5455. https://doi.org/10.1021/acs.iecr.9b07066.
(13) Bartnikowski, M.; Dargaville, T. R.; Ivanovski, S.; Hutmacher, D. W. Degradation Mechanisms of Polycaprolactone in the Context of Chemistry, Geometry and Environment. *Prog. Polym. Sci.* **2019**, *96*, 1–20. https://doi.org/10.1016/j.progpolymsci.2019.05.004.
(14) Coombes, A. G. A.; Rizzi, S. C.; Williamson, M.; Barralet, J. E.; Downes, S.; Wallace, W. A. Precipitation Casting of Polycaprolactone for Applications in Tissue Engineering and Drug Delivery. *Biomaterials* **2004**, *25* (2), 315–325. https://doi.org/10.1016/S0142-9612(03)00535-0.
(15) Militkỳ, J. Tensile Failure of Polyester Fibers. In *Handbook of properties of textile and technical fibres*; Elsevier, 2018; pp 421–514.
(16) Schanda, J. CIE Colorimetry. *Color. Underst. CIE Syst.* **2007**, *3*, 25–78.
(17) Hochberg, J. D.; Wirth, D. M.; Spiaggia, G.; Shah, P.; Rothen-Rutishauser, B.; Petri-Fink, A.; Pokorski, J. K. High-Throughput Manufacturing of Antibacterial Nanofibers by Melt Coextrusion and Post-Processing Surface-Initiated Atom Transfer Radical Polymerization. *ACS Appl. Polym. Mater.* **2022**, *4* (1), 260–269. https://doi.org/10.1021/acsapm.1c01264.
(18) Hochberg, J. D.; Wirth, D. M.; Pokorski, J. K. PET-RAFT to Expand the Surface-Modification Chemistry of Melt Coextruded Nanofibers. *Polym. Chem.* **2023**, *14* (9), 1054–1063. https://doi.org/10.1039/D2PY01389D.

(19) Hochberg, J. D.; Wirth, D. M.; Pokorski, J. K. Surface-Modified Melt Coextruded Nanofibers Enhance Blood Clotting In Vitro. *Macromol. Biosci.* **2022**, *22* (12), 2200292. https://doi.org/10.1002/mabi.202200292.

(20) Crescenzi, V.; Manzini, G.; Calzolari, G.; Borri, C. Thermodynamics of Fusion of Poly-β-Propiolactone and Poly-ϵ-Caprolactone. Comparative Analysis of the Melting of Aliphatic Polylactone and Polyester Chains. *Eur. Polym. J.* **1972**, *8* (3), 449–463. https://doi.org/10.1016/0014-3057(72)90109-7.

(21) Jordan, A. M.; Korley, L. T. J. Toward a Tunable Fibrous Scaffold: Structural Development during Uniaxial Drawing of Coextruded Poly(ε-Caprolactone) Fibers. *Macromolecules* **2015**, *48* (8), 2614–2627. https://doi.org/10.1021/acs.macromol.5b00370.

(22) Wang, S. S.; Magnusson, R. Theory and Applications of Guided-Mode Resonance Filters. *Appl. Opt.* **1993**, *32* (14), 2606–2613. https://doi.org/10.1364/AO.32.002606.

(23) Jang, J.; Badloe, T.; Sim, Y. C.; Yang, Y.; Mun, J.; Lee, T.; Cho, Y.-H.; Rho, J. Full and Gradient Structural Colouration by Lattice Amplified Gallium Nitride Mie-Resonators. *Nanoscale* **2020**, *12* (41), 21392–21400. https://doi.org/10.1039/D0NR05624C.

(24) Lu, Q.; Shu, F.-J.; Zou, C.-L. Extremely Local Electric Field Enhancement and Light Confinement in Dielectric Waveguide. *IEEE Photonics Technol. Lett.* **2014**, *26* (14), 1426–1429. https://doi.org/10.1109/LPT.2014.2322595.

(25) Ellenbogen, T.; Seo, K.; Crozier, K. B. Chromatic Plasmonic Polarizers for Active Visible Color Filtering and Polarimetry. *Nano Lett.* **2012**, *12* (2), 1026–1031. https://doi.org/10.1021/nl204257g.

(26) Keshavarz Hedayati, M.; Elbahri, M. Review of Metasurface Plasmonic Structural Color. *Plasmonics* **2017**, *12* (5), 1463–1479. https://doi.org/10.1007/s11468-016-0407-y.

(27) Kristensen, A.; Yang, J. K. W.; Bozhevolnyi, S. I.; Link, S.; Nordlander, P.; Halas, N. J.; Mortensen, N. A. Plasmonic Colour Generation. *Nat. Rev. Mater.* **2016**, *2* (1), 1–14. https://doi.org/10.1038/natrevmats.2016.88.

(28) Qin, J.; Jiang, S.; Wang, Z.; Cheng, X.; Li, B.; Shi, Y.; Tsai, D. P.; Liu, A. Q.; Huang, W.; Zhu, W. Metasurface Micro/Nano-Optical Sensors: Principles and Applications. *ACS Nano* **2022**, *16* (8), 11598–11618. https://doi.org/10.1021/acsnano.2c03310.

(29) Tan, S. J.; Zhang, L.; Zhu, D.; Goh, X. M.; Wang, Y. M.; Kumar, K.; Qiu, C.-W.; Yang, J. K. W. Plasmonic Color Palettes for Photorealistic Printing with Aluminum Nanostructures. *Nano Lett.* **2014**, *14* (7), 4023–4029. https://doi.org/10.1021/nl501460x.

(30) Yang, W.; Xiao, S.; Song, Q.; Liu, Y.; Wu, Y.; Wang, S.; Yu, J.; Han, J.; Tsai, D.-P. All-Dielectric Metasurface for High-Performance Structural Color. *Nat. Commun.* **2020**, *11* (1), 1864. https://doi.org/10.1038/s41467-020-15773-0.

(31) Proust, J.; Bedu, F.; Gallas, B.; Ozerov, I.; Bonod, N. All-Dielectric Colored Metasurfaces with Silicon Mie Resonators. *ACS Nano* **2016**, *10* (8), 7761–7767. https://doi.org/10.1021/acsnano.6b03207.

(32) Cao, L.; Fan, P.; Barnard, E. S.; Brown, A. M.; Brongersma, M. L. Tuning the Color of Silicon Nanostructures. *Nano Lett.* **2010**, *10* (7), 2649–2654. https://doi.org/10.1021/nl1013794.

(33) Brainard, D. H. Color Appearance and Color Difference Specification. *Sci. Color* **2003**, *2* (191–216), 5.

(34) Robertson, A. R. The CIE 1976 Color-Difference Formulae. *Color Res. Appl.* **1977**, *2* (1), 7–11. https://doi.org/10.1002/j.1520-6378.1977.tb00104.x.

(35) Ul-Hamid, A. *A Beginners' Guide to Scanning Electron Microscopy*; Springer International Publishing: Cham, 2018. https://doi.org/10.1007/978-3-319-98482-7.

(36) Ioannidi, E.; Risbo, J.; Aarøe, E.; van den Berg, F. W. J. Thermal Analysis of Dark Chocolate with Differential Scanning Calorimetry—Limitations in the Quantitative Evaluation of the Crystalline State. *Food Anal. Methods* **2021**, *14* (12), 2556–2568. https://doi.org/10.1007/s12161-021-02073-6.

(37) Schick, C. Differential Scanning Calorimetry (DSC) of Semicrystalline Polymers. *Anal. Bioanal. Chem.* **2009**, *395* (6), 1589–1611. https://doi.org/10.1007/s00216-009-3169-y.

(38) Hecht, E. *Optics*; Pearson Education: USA, 2017.

(39) Lin, Y.; Bilotti, E.; Bastiaansen, C. W. M.; Peijs, T. Transparent Semi-Crystalline Polymeric Materials and Their Nanocomposites: A Review. *Polym. Eng. Sci.* **2020**, *60* (10), 2351–2376. https://doi.org/10.1002/pen.25489.

(40) Cai, T.; Zhang, H.; Guo, Q.; Shao, H.; Hu, X. Structure and Properties of Cellulose Fibers from Ionic Liquids. *J. Appl. Polym. Sci.* **2010**, *115* (2), 1047–1053. https://doi.org/10.1002/app.31081.

(41) Himmler, M.; Schubert, D. W.; Fuchsluger, T. A. Examining the Transmission of Visible Light through Electrospun Nanofibrous PCL Scaffolds for Corneal Tissue Engineering. *Nanomaterials* **2021**, *11* (12), 3191. https://doi.org/10.3390/nano11123191.

(42) Haddadin, Z.; Khan, S.; Poulikakos, L. V. Cutting Corners to Suppress High-Order Modes in Mie Resonator Arrays. *ACS Photonics* **2024**, *11* (1), 187–195. https://doi.org/10.1021/acsphotonics.3c01270.